\documentclass[aps,pra,twocolumn,showpacs,superscriptaddress,reprin,nofootinbib]{revtex4-1}
\usepackage{epsfig}  
\usepackage[colorlinks,breaklinks]{hyperref}
\usepackage{nicefrac}
\usepackage{bm}
\usepackage{dcolumn}
\usepackage{graphics}
\usepackage{amsmath}
\usepackage{amssymb}
\usepackage{color}
\newcommand{\bra}{\langle}
\newcommand{\ket}{\rangle}

\newcommand{\be}{\begin{equation}}
\newcommand{\ee}{\end{equation}}
\newcommand{\bea}{\begin{align}}
\newcommand{\eea}{\end{align}}
\newcommand{\beqa}{\begin{eqnarray}}
\newcommand{\eeqa}{\end{eqnarray}}
\newcommand{\nablavec}{\ensuremath{\boldsymbol{\nabla}}}
\newcommand{\rvec}{\ensuremath{\boldsymbol{r}}}
\newcommand{\qvec}{\ensuremath{\boldsymbol{q}}}
\newcommand{\pvec}{\ensuremath{\boldsymbol{p}}}
\newcommand{\evec}{\ensuremath{\boldsymbol{\eta}}}

\newcommand{\vdW}{\mathrm{vdW}}

\newcommand{\simge}{\hspace*{0.2em}\raisebox{0.5ex}{$>$}
     \hspace{-0.8em}\raisebox{-0.3em}{$\sim$}\hspace*{0.2em}}
\newcommand{\simle}{\hspace*{0.2em}\raisebox{0.5ex}{$<$}
     \hspace{-0.8em}\raisebox{-0.3em}{$\sim$}\hspace*{0.2em}}

\begin{document}

\title{Effective field theory for few-boson systems}

\date{\today}

\author{Betzalel Bazak}
\email{bazak@ipno.in2p3.fr}
\affiliation{Institut de Physique Nucl\'eaire, CNRS-IN2P3, 
Univ. Paris-Sud, Universit\'e Paris-Saclay, 91406 Orsay, France}

\author{Moti Eliyahu}
\affiliation{The Racah Institute of Physics, The Hebrew University, 
Jerusalem 9190401, Israel}

\author{Ubirajara van Kolck}
\email{vankolck@ipno.in2p3.fr}
\affiliation{Institut de Physique Nucl\'eaire, CNRS-IN2P3, 
Univ. Paris-Sud, Universit\'e Paris-Saclay, 91406 Orsay, France}
\affiliation{Department of Physics, University of Arizona, Tucson, 
AZ 85721, USA}

\begin{abstract}
  We study universal bosonic few-body systems within the framework of effective
  field theory at leading order (LO).
  We calculate binding energies of systems of up to six particles
  and the atom-dimer scattering length.
  Convergence to the limit of zero-range two- and
  three-body interactions is shown, indicating that no additional 
  few-body interactions need to be introduced at LO.
  Generalizations of the Tjon line are constructed, showing correlations
  between few-body binding energies and the binding energy
  of the trimer, for a given dimer energy.
  As a specific example, we implement our theory for $^4$He atomic systems
  and show that the results are in surprisingly good agreement with those of 
  sophisticated $^4$He-$^4$He potentials. 
  Potential implications for the convergence of the EFT expansion are discussed.
\end{abstract}


\maketitle

\section{Introduction}
Universal systems are not sensitive to the details of their microscopic
physics. 
To a desired accuracy, their properties are frequently governed by
a much more restricted set of parameters than the ``underlying'' theory
of the microscopic dynamics.
Effective field theory (EFT) is the framework to 
formulate universal systems in a systematic expansion in a ratio
of small parameters so that at lowest orders
only a few parameters appear.
In this paper we discuss the properties of universal few-boson systems
from the perspective of EFT.

Consider particles interacting through 
an interaction with range $R$.
The effective range expansion (ERE) describes the 
two-body scattering process at relative 
wavenumber $k\ll 1/R$
through a power series in $k^2$,
\be
k \cot \delta (k) = 
-\frac{1}{a_2} + \frac{r_2 }{2} k^2 +
\mathcal{O}\left(k^4 \right),
\label{ERE}
\ee
where 
$\delta(k)$ is the $s$-wave phase shift,
$a_2$ is the scattering length, $r_2$ is the effective range,
and further parameters appear in the higher-power terms.
Typically, all ERE parameters have a size set by $R$, for example 
$|a_2| \approx |r_2| \approx R$. 
A type of universality occurs when the system is 
fine-tuned such that the scattering length is large compared to the other
ERE parameters, 
$|a_2| \gg |r_2| \approx R$. 
In this case the two-body system of reduced mass $\mu$
has a shallow real or virtual bound state
at a binding energy
$B_2\approx \hbar^2/(2\mu a_2^2) \ll \hbar^2/(2\mu R^2)$,
with corrections on the order of $R/|a_2|$.
As first shown by Efimov \cite{Efimov:1970zz}, 
a number of shallow bound states appear 
when more particles are present.
The properties of these systems are characterized by 
only a few parameters; for a review, see Ref. \cite{BraHam06}.

It is indeed common in physics to have an underlying theory, valid at a 
momentum scale $M_{hi}$, while we are interested in processes 
occurring at typical momenta $Q$ which are comparable to a much lower 
momentum scale $M_{lo}$, $Q \approx M_{lo} \ll M_{hi}$.
For example, nuclear structure involves momenta that are much smaller
than the typical momentum scale of QCD, $M_{QCD} \approx 1$ GeV/$c$. 
In this case, the range of the two-nucleon force
$R \approx \hbar / m_\pi c \simeq 1.4$ fm, which is set by the pion mass
$m_\pi$, is thought to be relevant for heavy nuclei.
However, the two-nucleon scattering lengths in both spin-singlet and 
spin-triplet $s$ waves, respectively 
$a_{2s}\simeq -23.4$ fm and $a_{2t}\simeq 5.42$ fm,
are larger in magnitude than $R$.
Nuclear few-body systems thus fall into the same universality
class as other systems characterized by large scattering lengths.

Another system in this class, which has even larger scale separation, is   
the $^4$He atomic system \cite{LuoGieGen96}. Here $a_2\approx 180 \, a_0$, 
where $a_0$ is the Bohr radius,
is much larger than the van der Waals radius, $R\approx 10 \, a_0$.
In addition, over the last two decades much attention has been devoted to
an artificial system where these scales can be tuned by hand,
namely ultracold atoms near a Feshbach resonance.
Here $a_2$ can be tuned to any value by changing an external magnetic
or electric field \cite{ChiGriJul10}.

Effective field theory allows us to focus on the low-momentum region in the
more general case
where there exist light particles (such as pions in nuclear physics)
that generate long-range forces,
which invalidate a straightforward ERE at two-body level.
In EFT one starts by constructing the most general effective Lagrangian
by integrating out the high-energy degrees of freedom, while
keeping the symmetries of the underlying theory.
The details of the underlying dynamics are contained in the
interaction strengths, called low-energy constants (LECs).
Through an estimate of the effects of interactions on observables,
called power counting, 
a controlled approximation for the system at hand is obtained.
Even when we are at such low energies that the ERE applies,
the EFT with $M_{hi}\sim \hbar/R$ allows us to account for deviations
from the zero-range limit in a systematic expansion in 
$M_{lo}/M_{hi}\sim R/|a_2|$,
not only for the two-body system but also for more-particle
bound states, as long as they are sufficiently shallow,
and for reactions involving these states.

The EFT that reproduces the ERE for any short-range force
was formulated at the two-body level in Ref. \cite{vanKolck:1998bw}.
It was shown that 
at leading order (LO) a single two-body delta-function interaction
appears which captures the information about $a_2$,
while at higher orders other ERE parameters enter through more-derivative 
contact forces.
After renormalization,
the two-body amplitude is equivalent not only to the
ERE, but also to Fermi's pseudopotential
and to an energy-dependent boundary condition at the origin.
The extension to three-boson systems was formulated 
at LO in Refs. \cite{BedHamKol99,BedHamKol99b,MohFurHam06}, where it was found
that a three-body, no-derivative contact force with a parameter 
$\Lambda_\star$ is needed for proper renormalization.
The three-body force is on a limit cycle and
$\Lambda_\star$ controls the position of the tower
of Efimov states, for example the binding
energy $B_3$ of the three-body ground state.
As $\Lambda_\star$ is varied at fixed $a_2$, correlations appear
among three-boson observables, such as the bosonic analog 
of the Phillips line \cite{Phillips:1968zze}, 
which relates the atom-dimer scattering length $a_3$ to $B_3$.
The successful, fully perturbative extension to the next two orders 
of the expansion was carried
out in Refs. \cite{Ji:2011qg} and \cite{Ji:2012nj}, the latter
explicitly 
demonstrating the appearance of a new three-body force, with a new parameter.
The minimum orders at which three-body forces of various types appear have
been determined in Ref. \cite{Griesshammer:2005ga}.

Two questions arise:
\begin{enumerate}
\item Are higher-body forces needed at LO to describe systems with more
  bodies?
  A first step in this direction was taken in
  Refs. \cite{PlaHamMei04,Hammer:2006ct} where the four-boson system was 
  shown to be properly renormalized, at least within the first cycle of
  the three-body force.
  As a consequence of the absence of a LO four-body force,
  correlations develop between four- and three-body observables
  such as the bosonic analog of the Tjon line \cite{Tjo75}, which relates
  the binding energy $B_4$ of the four-body ground state to $B_3$.
  Although no higher-order calculation of this system exists,
  it is inevitable that at some order a four-body force will appear
  and introduce a new scale. 
  The importance of this new scale has been the subject of debate
  \cite{Hadizadeh:2011qj,Frederico:2012az}.

\item What is the regime of validity of the EFT as the number
  of particles increases? 
  As shown in Ref. \cite{Hammer:2006ct}, a three-body Efimov state spawns
  two four-body states, one barely more bound and
  another much more bound than the three-body parent. 
  Model calculations \cite{Ste10,GatKieViv11,Ste11,Gattobigio:2012tk} 
  indicate that the same phenomenon
  repeats as the number of bodies is increased further.
  Thus, one expects deeper and deeper states. 
  A measure of the particle binding momentum for a system of $N$ identical 
  particles of mass $m$ is
  \begin{equation}
  Q_N=\sqrt{\frac{2m B_N}{N}}.
  \label{QA}
  \end{equation}
  The EFT expansion in $R/|a_2|$ includes, for few-body systems, 
  also an expansion in $Q_NR$, 
  which increases as the binding energy per particle, $B_N/N$, increases. 

\end{enumerate}

In order to make progress in answering these questions,
here we build the EFT for few-boson systems with a large two-body scattering 
length, and study such systems with up to six particles at LO.
One calculation for $N>4$ particles exists
in short-range EFT: the binding energy of $N=6$ nucleons at
LO \cite{SteBarKol07} is consistent with both the absence of LO higher-body
forces and EFT convergence, but is not conclusive
\footnote{Note that finite-range potential models, 
such as those employed in the $N>4$ calculations of 
Refs. \cite{Ste10,GatKieViv11,Ste11,Gattobigio:2012tk,YanBlu15},
resemble an EFT with a finite regulator, and most of their results
are likely to be compatible with EFT.
However, in an EFT regulator insensitivity is needed for model independence,
while physical range effects are subleading and should be treated 
in perturbation theory \cite{vanKolck:1998bw,Ji:2011qg}.}.

We extend the LO results to up to six bosons
by solving the Schr\"odinger equation 
with the stochastic variational method (SVM) \cite{SuzVar98}.
With an eye to future extensions employing Monte Carlo methods,
we use a local Gaussian regulator.
Our approach is valid for any bosonic system
with large scattering length,
including those close to the unitary limit $|a_2|\to \infty$.
In order to be concrete, we apply the EFT to $^4$He atoms,
motivated by the recent experimental verification of the excited Efimov
state in this system \cite{Kun15}.
Helium systems have been studied extensively with various 
$^4$He-$^4$He potentials, see for example Refs.
\cite{NakAkaTanLim78,NakLimAkaTan79,LimNakAkaTan80,AziSla91,JanAzi95,Lew97,
BluGre00,Rou03,KolMotSan04,LazCar06,PrzCenKom10,HiyKam12a,Del15}.
In EFT, the three-$^4$He system has been studied up to N$^2$LO
\cite{BedHamKol99,BedHamKol99b,BraHam03,Ji:2012nj}
using a formulation based on a dimer auxiliary field and on a 
sharp-momentum regularization
with two- and three-body parameters $\Lambda_2\gg \Lambda_3$.
The four-$^4$He system was calculated at LO \cite{PlaHamMei04}
by solving the Yakubovsky equations with a non-local Gaussian regulator.

For $N\le 4$ our results are consistent with previous works.
For a suitable two-body interaction strength a shallow dimer exists
with a spatial extent that is about
10 times larger than the range associated with the van der Waals interaction
between its constituents. 
Two trimers are bound:
one shallow, close to the dimer, and one much deeper, about 50 times more 
bound. 
The ground-state trimer, in turn,
is followed by two tetramer states, one close to the trimer and one deeper.
When the excited trimer binding energy $B_3^*$ is used to fix the three-body
force parameter, the ground-state trimer and tetramer binding energies 
converge in the zero-range limit.
Both trimer and tetramer \cite{NakAkaTanLim78} ground-state energies
are correlated with the excited trimer energy. 

For $N=5,6$ we find similar convergence of binding energies 
with increasing cutoff parameter,
suggesting that no higher-body forces are needed in LO.
As a consequence, we can construct generalized Tjon lines 
where the binding energies beyond $N=4$ are correlated with
$B_3$ at fixed two-body input.
Such lines have been constructed before by combining
results from different phenomenological two-body potentials
\cite{NakLimAkaTan79,LimNakAkaTan80}; here, these lines
arise from the continuous 
variation of the single LO three-body parameter $\Lambda_\star$.
Deeper states are smaller, and therefore finite-range corrections are
expected to be more important for them. 
The ground-state energies grow to about 16 times
the trimer energy at $N=6$, but we will see that our results still 
agree with potential models within about 15\%.
Although
we cannot offer a conclusive answer without higher-order calculations,
our LO calculations are very encouraging for the convergence of the EFT
in many-body systems. They also provide a baseline for future comparisons
with higher-order results.

Our paper is organized as follows.
In Sec. \ref{sec:theory} we introduce the LO EFT, and its regularization
and renormalization.
We also discuss the two inputs needed to fix the two LO parameters.
In Sec. \ref{sec:methods} the numerical method for the solution
of the Schr\"odinger equation is presented.
Our results are given in Sec. \ref{sec:results}, while
Sec. \ref{sec:summary} summarizes our findings and some of their implications.

\section{Theory}
\label{sec:theory}

A system of spinless bosons of mass $m$ interacting via a short-range
force can be described by the Lagrangian density
\be
\mathcal L = \psi^\dagger \left(i\partial_0 
+ \frac{\nablavec^2}{2m} \right) \psi
- \frac{\tilde C^{(0)}}{2} (\psi^\dagger \psi)^2 
- \frac{\tilde D^{(0)}}{6}  (\psi^\dagger \psi)^3 + \ldots 
\label{Leff}
\ee
where $\psi$ is the bosonic field operator,
$\tilde C^{(0)}$ and $\tilde D^{(0)}$ are low-energy constants,
and ``$\ldots$'' stand for terms with more fields and/or more derivatives,
which are subdominant. Since this work is
focused on the leading order, these terms will be neglected in the following. 
For simplicity we use units where $\hbar=c=1$.

The interactions in Eq. \eqref{Leff} are $delta$ functions in coordinate space.
For singular interactions, the solution of the Schr\"odinger equation
requires regularization, that is, the introduction
of a function that suppresses momenta above a cutoff $\Lambda$.
This is achieved 
by smearing the delta function over distances $\sim \Lambda^{-1}$.
Observables are independent of the arbitrary value of $\Lambda$
(renormalization)
if the LECs have specific dependences on $\Lambda$.
In the following we use dimensionless LECs by writing
\be
\tilde C^{(0)}(\Lambda)= \frac{4\pi}{m\Lambda}C^{(0)}(\Lambda),
\quad
\tilde D^{(0)}(\Lambda)= \frac{(4\pi)^2}{m\Lambda^4}D^{(0)}(\Lambda).
\label{dimlessLECs}
\ee
When the EFT is truncated at given order, observables acquire a residual
cutoff dependence ${\cal O}(Q/\Lambda)$, which can be made arbitrarily small
by increasing $\Lambda$. However, the truncation of the Lagrangian
induces relative errors of ${\cal O}(Q/M_{hi})$. Thus, a variation
of the cutoff from $M_{hi}$ to much larger values gives an estimate of
the theoretical error. (For values of $\Lambda$ below $M_{hi}$,
the error is dominated by the regularization error.)
We discuss our specific regularization and renormalization procedures next.

\subsection{Two-body sector}
\label{sec:theory2}

At low energies, and since $|a_2|$ is much larger than 
the range of the potential,
physics cannot be sensitive to the short-distance details of the potential.
Therefore the potential between the particles can be represented
by a contact potential, $V = \tilde C^{(0)} \delta(\rvec)$.
To see the need for regularization, one can solve 
the momentum-space Schr\"odinger equation for the two-body bound state,
\be
\frac{p^2}{m} \phi(p) + \tilde C^{(0)} \int \frac{d^3p'}{(2\pi)^3}\phi(p')
= -B_2 \phi(p)
\ee
where $B_2$ is the dimer binding energy and
$\phi(p)$ is the momentum-space radial wave function.
Solving for $\phi(p)$,
the coupling constant $\tilde C^{(0)}$ has to satisfy 
\be
\frac{1}{\tilde C^{(0)}}=-\int \frac{d^3p'}{(2\pi)^3} \frac{1}{p'^2/m + B_2}.
\ee
However, since the integral on the right-hand side diverges,
no solution exists for any finite $\tilde C^{(0)}$.

If we denote the incoming (outgoing) relative momenta in two-body scattering
by  $\pvec$ ($\pvec '$),
one can regularize the integral by introducing cutoff functions
$f_\Lambda(\pvec)f_\Lambda(\pvec')$, where $\Lambda$ is the cutoff parameter.
Using a Gaussian function, $f_\Lambda(x)=\exp(-x^2/\Lambda^2)$, 
the equation for $\tilde C^{(0)}$ is modified to  
\be
\frac{1}{\tilde C^{(0)}(\Lambda)}=-\int \frac{d^3p'}{(2\pi)^3}
\frac{\exp(-2p'^2/\Lambda^2)}{p'^2/m + B_2},
\ee
which converges. The solution can be expanded in powers of $Q_2/\Lambda$,
where $Q_2$ is defined in Eq. \eqref{QA}:
\be
\tilde C^{(0)}(\Lambda)=-\frac{4\pi}{m\Lambda}
\sqrt{2\pi}\left[1 + \sqrt{2\pi}\frac{Q_2}{\Lambda} +
\mathcal{O} \left(\frac{Q_2^2}{\Lambda^2}\right) \right].
\label{C0sep}
\ee
Therefore, to keep the dimer binding energy fixed, the coupling constant has
to ``run'' with the cutoff parameter $\Lambda$. 
For another regulator
the dependence of $\tilde C^{(0)}$ on $\Lambda$ will not, in general,
be given by Eq. \eqref{C0sep}, but it will still be such as
to reproduce $B_2$.
Any physical observable has to be independent of 
$\Lambda$ and of the regularization method used,
within the error generated by the neglect of higher orders.

In coordinate space the contact potential is now smeared over a
range $\sim 1/\Lambda$.
For the non local Gaussian regulator the result is 
a non-local potential \cite{MohFurHam06},
\be
\bra \rvec | V | \rvec' \ket = \tilde C^{(0)}(\Lambda) 
\, \delta_\Lambda(\rvec) \, \delta_\Lambda(\rvec' ),
\ee
where
\be
\delta_\Lambda(\rvec) \equiv
\frac{\Lambda^3}{8\pi^{3/2}} \exp\left(-\Lambda^2 \rvec^2/4\right)
\label{gauss}
\ee
satisfies $\lim_{\Lambda\to\infty}\delta_\Lambda(\rvec)=\delta(\rvec)$.

To get a local potential we use a regularization $f_\Lambda(\qvec)$
on the momentum transfer $\qvec=\pvec-\pvec'$, yielding instead
\be
\bra \rvec | V | \rvec' \ket = \tilde C^{(0)}(\Lambda) \, \delta_\Lambda(\rvec) 
\, \delta(\rvec-\rvec').
\ee
Defining the dimensionless LEC according to Eq. \eqref{dimlessLECs}
and summing over all pairs, the two-body interaction is
\be
V_{2b} = \frac{4\pi}{m\Lambda}
C^{(0)}(\Lambda) \sum_{i<j} \delta_\Lambda(\rvec_{ij}),
\label{Heff}
\ee
where $\rvec_{ij}$ is the position of particle $i$ with respect to particle $j$.
We will employ the Gaussian regularization \eqref{gauss}.
To renormalize $C^{(0)}(\Lambda)$ one has to choose an observable to
be fixed to its physical value. 
The parameter $C^{(0)}(\Lambda)$ is then, in fact, a function
of the dimensionless ratio $Q_2/\Lambda$.
However, for a local regulator the running of $C^{(0)}(\Lambda)$
cannot be obtained analytically; we discuss it in 
Sec. \ref {sec:results2} below.

\subsection{Three-body sector}
\label{sec:theory3}

Although the two-body system is suitably renormalized,
the trimer ground-state binding energy $B_3$ is found to
be strongly dependent on the cutoff \cite{BedHamKol99,BedHamKol99b}, 
$B_3 \propto \Lambda^2/m$, a result first obtained by Thomas
\cite{Thomas:1935zz}.
As $\Lambda$ increases, other bound states appear in the spectrum.
Since $\Lambda$ is not a physical parameter,
a three-body contact interaction has to be introduced
at this order for renormalization \cite{BedHamKol99,BedHamKol99b}.
  
In Refs. \cite{BedHamKol99,BedHamKol99b}, 
the two-body amplitude in the large-cutoff limit was
used as input to solve the three-body problem,
and a sharp cutoff function $f_\Lambda(x) = \theta (1-x/\Lambda)$ was introduced
in the resulting integral equation.
The running of the three-body LEC was found to be
\be
\tilde D^{(0)}(\Lambda)\propto \frac{(4\pi)^2}{m\Lambda^4}
\frac{\sin \left(s_0 \ln (\Lambda/\Lambda_\star)-\arctan s_0^{-1}\right)}
     {\sin \left(s_0 \ln (\Lambda/\Lambda_\star)+\arctan s_0^{-1}\right)},
\label{limcyc}
\ee
where $s_0\simeq 1.00624$
and $\Lambda_\star$ is a parameter determined by one three-body datum.
A similar structure was found with the non-local Gaussian regulator
\cite{MohFurHam06,PlaHamMei04}.
This log-periodic structure has divergences,
signaling the appearance of deep bound states. 

Here we employ the same local Gaussian regulator as in the two-body system. 
With the dimensionless LEC introduced in Eq. \eqref{dimlessLECs}
the three-body interaction can be written as a cyclic permutation
of triplets,
\be
V_{3b} = \frac{(4\pi)^2}{m\Lambda^4}
D^{(0)}(\Lambda) \sum_{i<j<k}\sum_{cyc} 
\delta_\Lambda(\rvec_{ij})\, \delta_\Lambda(\rvec_{jk}).
\ee
The parameter $D^{(0)}(\Lambda)$ now depends not only on $Q_2/\Lambda$
but also on $\Lambda_\star/\Lambda$.
We discuss the solution of the three-boson system with our local regulator in
Sec. \ref{sec:results3}.

\subsection{$^4$He atoms}

The previous arguments and most of the issues we address below
concern any universal bosonic system,
and also apply to fermions with three or more states;
for a review, see Ref. \cite{Bedaque:2002mn}.
(For fermions with less than three states, there is no LO three-body force.)
To be definite, we focus on $^4$He atomic systems when presenting numerical
results.

The $^4$He atomic system was the subject of much study, both theoretical
and experimental.
The length scales associated with the two-body system are
the main ERE parameters, $a_2$ and $r_2$, as well as the van der Waals length,
defined as $r_{\vdW}=(m C_6)^{1/4}$, where $C_6$ characterizes the potential
tail $-C_6/r^6$.
In Table \ref{tbl:HeLengths} we present these length scales for $^4$He atoms
as obtained by two modern $^4$He-$^4$He potentials, 
LM2M2 \cite{AziSla91} and PCKLJS \cite{PrzCenKom10}.
As expected in this universality class, $a_2\gg r_2\sim r_{\vdW}$. 

\begin{table}
\begin{center}
  \caption{Length scales (in \AA) for two $^4$He atoms, deduced from two
    modern $^4$He-$^4$He potentials, LM2M2 and PCKLJS. 
    The scattering length $a_2$ and effective range $r_2$ are from
    Refs. \cite{JanAzi95,KolMotSan04,PrzCenKom10}, while 
    the van der Waals length $r_{\vdW}$ is evaluated from the value 
    of the van der Waals coefficient 
    $C_6$ calculated in Refs. \cite{YanBabDal96,ZhaYanVri06}.}
\label{tbl:HeLengths}
\vspace{0.3cm}
{\renewcommand{\arraystretch}{1.25}%
\begin{tabular}
{c | c@{\hspace{5mm}} c}
\hline\hline 
         & LM2M2  & PCKLJS    \\ 
\hline
$a_2$    & 100.23 & 90.42(92) \\ 
$r_2$    & 7.326  & 7.27      \\ 
$r_{\vdW}$ & 5.378  & 5.378     \\ 
\hline\hline
\end{tabular}}
\end{center}
\end{table}

In Table \ref{tbl:HeEn} we summarize the energies of $^4$He clusters,
as calculated from the same two potentials in Ref. \cite{HiyKam12a}.
We show in Table \ref{tbl:HeEn} also the results extracted from experimental
data.
The dimer binding energy $B_2$ is calculated in Ref. \cite{CenPrzKom12}
from the measurement of the average separation,
$\bra r \ket = 52(4)$ \AA~\cite{GriSchToe00}.
Very recently, a value of $1.76(15)$ mK was measured for the dimer
energy \cite{Zel16}. The difference between excited-trimer and dimer
energies, $B_3^*-B_2$, was also measured recently \cite{Kun15}.

\begin{table}
\begin{center}
\caption{Binding energies (in mK) of $^4$He clusters. $B_N^{(*)}$ denotes the
binding energy of the $N$-body ground (lowest excited) state.
The first two columns show results \cite{HiyKam12a}
for two modern $^4$He-$^4$He potentials.
The last column displays experimental results \cite{GriSchToe00,Zel16,Kun15}, 
as described in the text.}
\label{tbl:HeEn}
\vspace{0.3cm}
{\renewcommand{\arraystretch}{1.25}%
\begin{tabular}
{c | c@{\hspace{5mm}} c@{\hspace{5mm}} c} 
\hline\hline 
          & LM2M2 & PCKLJS & experiment         \\ 
\hline
$B_2$     & 1.3094 & 1.6154 & $1.3^{+0.25}_{-0.19}$,\, 1.76(15) \\ 
$B_3^*$   & 2.2779 & 2.6502                      \\ 
$B_3^*-B_2$& 0.9685 & 1.0348 & 0.98(2)            \\ 
$B_3$     & 126.50 & 131.84                      \\ 
$B_4^*$   & 127.42 & 132.70                      \\ %
$B_4$     & 559.22 & 573.90                      \\ %
\hline\hline
\end{tabular}}
\end{center}
\end{table}

Renormalization at the two- and three-body levels requires the input
of two observables to determine the running of $C^{(0)}(\Lambda)$
and $D^{(0)}(\Lambda)$.
We fit $C^{(0)}$ to the dimer binding energy $B_2$,
and $D^{(0)}$ to the excited-trimer binding energy $B_3^*$.
The excited trimer is closer to the three-boson threshold than the ground 
trimer and thus, presumably, afflicted by smaller errors 
stemming from the EFT truncation.
This induces a dependence of the three-body LEC $D^{(0)}$ 
on the ratio $Q_3^*/\Lambda$, where $Q_3^*=\sqrt{2m B_3^*/3}$
is the binding momentum of the trimer excited state, which
is related to the $\Lambda_\star$ introduced in Sec. \ref{sec:theory3}.

One would like to use the experimental data to calibrate the EFT.
However, there are two problems.
First, the experimental situation is not entirely clear, as can be
seen from the discrepancy in the dimer binding energies in Table \ref{tbl:HeEn}.
Second, there are no other experimental numbers to compare our predictions with,
even if we neglect the spread in dimer binding energies.
We use instead values calculated from the two modern potentials
LM2M2 and PCKLJS, which give $B_2$ and $B_3^*-B_2$ in fair agreement
with data and provide a reasonable representation of the data spread.
For these potentials a number of other observables have been calculated,
comparison with which offers an assessment of the success of the EFT
at LO. Needless to say, as the experimental situation improves
our calculations can be repeated with experimental input,
short-circuiting the need for potential models.

From the values of $B_2$ and $B_3^*$, the input binding momenta are
$Q_2\simeq 0.0055 \, a_0^{-1}$ and $Q_3^*\simeq 0.0059 \, a_0^{-1}$
for the LM2M2 potential, and
$Q_2\simeq 0.0061 \, a_0^{-1}$ and $Q_3^*\simeq 0.0064 \, a_0^{-1}$
for the PCKLJS potential.
With this input,
LO predictions can be made for systems with more particles.
Our method of solution of the Schr\"odinger equation is described in the
next section. 
Calculations are performed at various cutoff values.
The residual cutoff dependence of an observable can be expanded
in a power series in $Q/\Lambda$, where $Q$ is
the typical momentum associated to the physical process where that
observable is measured.
After fitting the coefficients of the power series
to the numerical results at cutoff values $\Lambda\simge 1/R$,
we can extrapolate to $\Lambda\to \infty$,
which corresponds to a zero-range potential.

However, in any physical situation the underlying interactions
have a finite range $R$, effects of which are accounted for
by higher-order EFT interactions.
A rough estimate of the LO relative error is $QR$,
where we take $1/R \sim 2/r_2\gg 1/a_2$
from Table \ref{tbl:HeLengths}.
An alternative error estimate is obtained from the residual
cutoff dependence.
If $Q$ is identified correctly, the coefficients of the
series in $Q/\Lambda$ are expected to
be of ${\cal O}(1)$. Then, by varying the cutoff from 
$\sim 1/R$ to $\infty$, we obtain an estimate of the error,
which should be comparable to the rougher $QR$ estimate.
Note that the two estimates can easily differ by a factor of ${\cal O}(1)$,
which is the expected size of the series coefficients.
Yet another way to gauge the size of higher-order effects is
to use different inputs that are supposed to be the same
at the order of interest, for example $1/a_2$ or $Q_2$ at LO.
Better estimates can be obtained once a few subleading orders have been
calculated, and one can be more confident about the magnitude
of the product $QR$ for a given observable.

\section{Method}
\label{sec:methods}

To solve the $N$-body Schr\"odinger equation, we use $N-1$ Jacobi vectors
$\evec_n$, $n=1,\ldots,N-1$, which we collect in a ``vector of vectors''
$\evec =(\evec_1,\evec_2 , \ldots, \evec_{N-1})^T$.
We expand the wavefunction on a correlated Gaussian basis,
\be
\psi(\evec)=
\sum_i c_i \, \hat {\mathcal S} \, \exp(-\frac{1}{2}\evec^T A_i \evec),
\ee
where 
$A_i $ is an $(N-1)\times(N-1)$ real, symmetric, and positive-definite matrix
and $c_i$ is a coefficient to be determined.
Here $\hat {\mathcal S}$ is the symmetrization operator
necessary to enforce bosonic symmetry.

One of the advantages of this basis is that the matrix elements
of the Hamiltonian as well as the overlap of two basis functions
can be calculated analytically in most cases.
Denoting by $|A\ket$ a basis vector with matrix $A$,
the overlap matrix element is
\be
\bra A | A' \ket = \left ( \frac{(2\pi)^{N-1}}{\det B} \right)^{3/2},
\ee
where $B=A+A'$.
The matrix element of the internal kinetic energy is
\be
\frac{\bra A | T_{int} | A' \ket}{\bra A | A' \ket} = 3 \, \mathrm{Tr} G ,
\ee
where $G = A B^{-1}A'\Pi$,  
with $\Pi_{nl}=(2\mu_n)^{-1}\delta_{nl}$ an $(N-1)\times(N-1)$ diagonal matrix
and $\mu_n$ the reduced mass corresponding to coordinate $\evec_n$.
The matrix elements of the potential involve the relative particle positions,
which we write as $\rvec_{ij} =\omega_{ij}^T \evec$.
For a fixed pair $ij$, $\omega_{ij}$ is an $N-1$-component vector. 
The two-body interaction matrix element is then
\be
\frac{\bra A | V_{2b} | A' \ket}{\bra A | A' \ket} = 
\frac{\Lambda^2 C^{(0)}(\Lambda)}{2 \sqrt{\pi} m}
\sum_{i<j} \left(1 + f_{ij} \frac{\Lambda^2}{2}\right)^{-3/2},
\ee
where $f_{ij}=\omega_{ij}^T B^{-1}\omega_{ij}$.
For the three-body interaction,
\beqa
\frac{\bra A | V_{3b} | A' \ket}{\bra A | A' \ket} &=&
\frac{\Lambda^2  D^{(0)}(\Lambda)}{4 \pi m}
\nonumber\\
&&\times \sum_{i<j<k} \sum_{cyc}
\left[\det\left(I+ F_{ijk}\frac{\Lambda^2}{2}\right) \right]^{-3/2},
\eeqa
where $I$ is the $2\times 2$ identity matrix
and $F_{ijk}=\Omega_{ijk}^T B^{-1} \Omega_{ijk}$ is a $2\times 2$ matrix,
with $\Omega_{ijk}=(\omega_{ik} \; \omega_{jk})$ a $2\times (N-1)$ matrix.

Another advantage of this basis is its flexibility:
Since we want $\Lambda^{-1} \simle R \ll a_2$
we need a large spread of basis functions, which
can be achieved within this basis.

To optimize our basis we use the
stochastic variational method (SVM) \cite{SuzVar98}.
To add a function to our basis, or to refine our basis by replacing
an exist basis function with a new one,
the elements of the matrix $A_i$ are chosen randomly one by one,
and the one which gives the lowest energy is taken.
According to the variational principle, upper bounds
for the ground and excited states are guaranteed.

\section{Results}
\label{sec:results}

Here we present the results of the LO EFT (Sec. \ref{sec:theory}) 
solved with the SVM (Sec. \ref{sec:methods}) for $N=2-6$ particles.

\subsection{Two bosons}
\label{sec:results2}

\begin{figure}
\begin{center}
\includegraphics[width=8.6 cm]{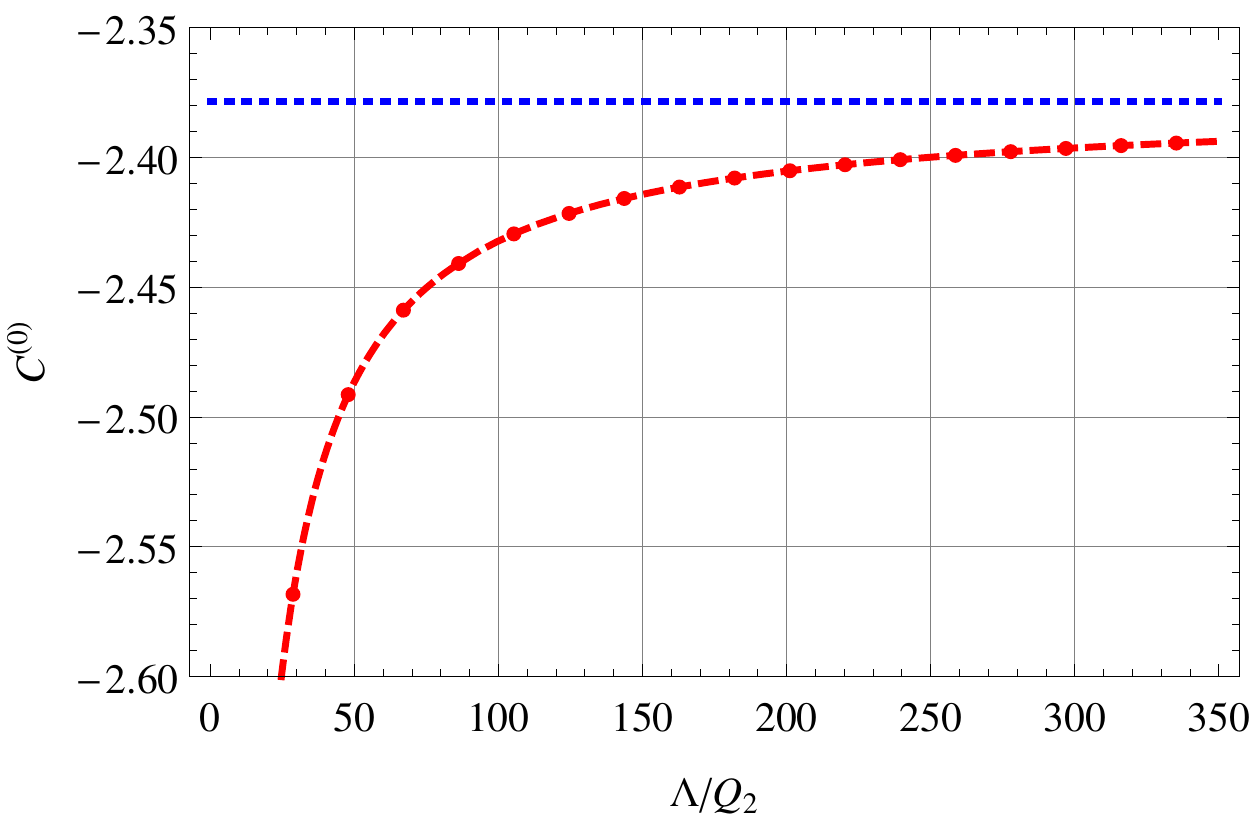}
\caption{\label{fig:C0} (Color online)
  The dimensionless two-body low-energy constant $C^{(0)}$ for various cutoff 
  values $\Lambda$, in units of the dimer binding momentum $Q_2$.
  Shown are the results for $^4$He atoms (red circles) and a fit with
  Eq. \eqref{C0nonsep} (red dashed line) which has the asymptotic value
  $C^{(0)}_\infty$ (blue dotted, horizontal line).}
\end{center}
\end{figure}

We solve the two-body Schr\"odinger equation and
demand that the dimer binding energy $B_2$ be reproduced
for any value of the cutoff.
This determines the values of the LEC $C^{(0)}(\Lambda)$, 
which is shown in Fig. \ref{fig:C0}.
The numerical results are well fitted by
\be
C^{(0)}(\Lambda) = C^{(0)}_\infty
\left [1+ \alpha \frac{Q_2}{\Lambda} +
\beta \left(\frac{Q_2}{\Lambda}\right)^2 +\ldots\right],
\label{C0nonsep}
\ee
where $C^{(0)}_\infty \simeq -2.379$, $\alpha \simeq 2.241$ and
$\beta \simeq 1.456$.
This curve is universal in the sense that at the same cutoff $\Lambda$
different input potentials, which differ in $Q_2$, correspond to points
lying on this curve.
The $Q_2$-independent
asymptotic value $C^{(0)}_\infty$ is also indicated in Fig. \ref{fig:C0}.

Equation \eqref{C0nonsep} can be compared with Eq. \eqref{C0sep},
which was also obtained with a Gaussian regulator, but a separable one.
In both cases, the expansion coefficients are
numbers of ${\cal O}(1)$, as expected from the fact that the
two-body binding momentum $Q_2$ provides the scale for cutoff variation
of the LEC.

At this order, the two-body amplitude gives rise to the 
ERE expansion \eqref{ERE} with $1/a_2 = Q_2$.
The other ERE parameters are cutoff dependent and vanish
as $\Lambda \to \infty$.
For example, the induced effective range is
\be
r_\mathrm{2}(\Lambda) = \frac{\gamma}{\Lambda}
\left[1+{\cal O}\left(\frac{Q_2}{\Lambda}\right)\right],
\label{r2}
\ee
where $\gamma \simeq 2.869$.
This is because the regularized two-body potential \eqref{Heff}
has a range $\sim\Lambda^{-1}$.
In the limit of $\Lambda\to\infty$ we reproduce, therefore, the
limit of a zero-range potential.
The $\Lambda^{-1}$ dependence in Eq. \eqref{r2} implies 
two-body corrections to the zero-range limit appear at 
next-to-leading order (NLO).
At this order a two-derivative delta-function potential must be introduced
in first-order distorted-wave Born approximation,
and its LEC adjusted to yield a finite effective range $r_2$
\cite{vanKolck:1998bw}.
At next-to-next-to-leading order
(N$^2$LO), corrections proportional to $r_2^2$ determine the 
two-body amplitude \cite{vanKolck:1998bw}.

\subsection{Three bosons}
\label{sec:results3}

\begin{figure}
\begin{center}
\includegraphics[width=8.6 cm]{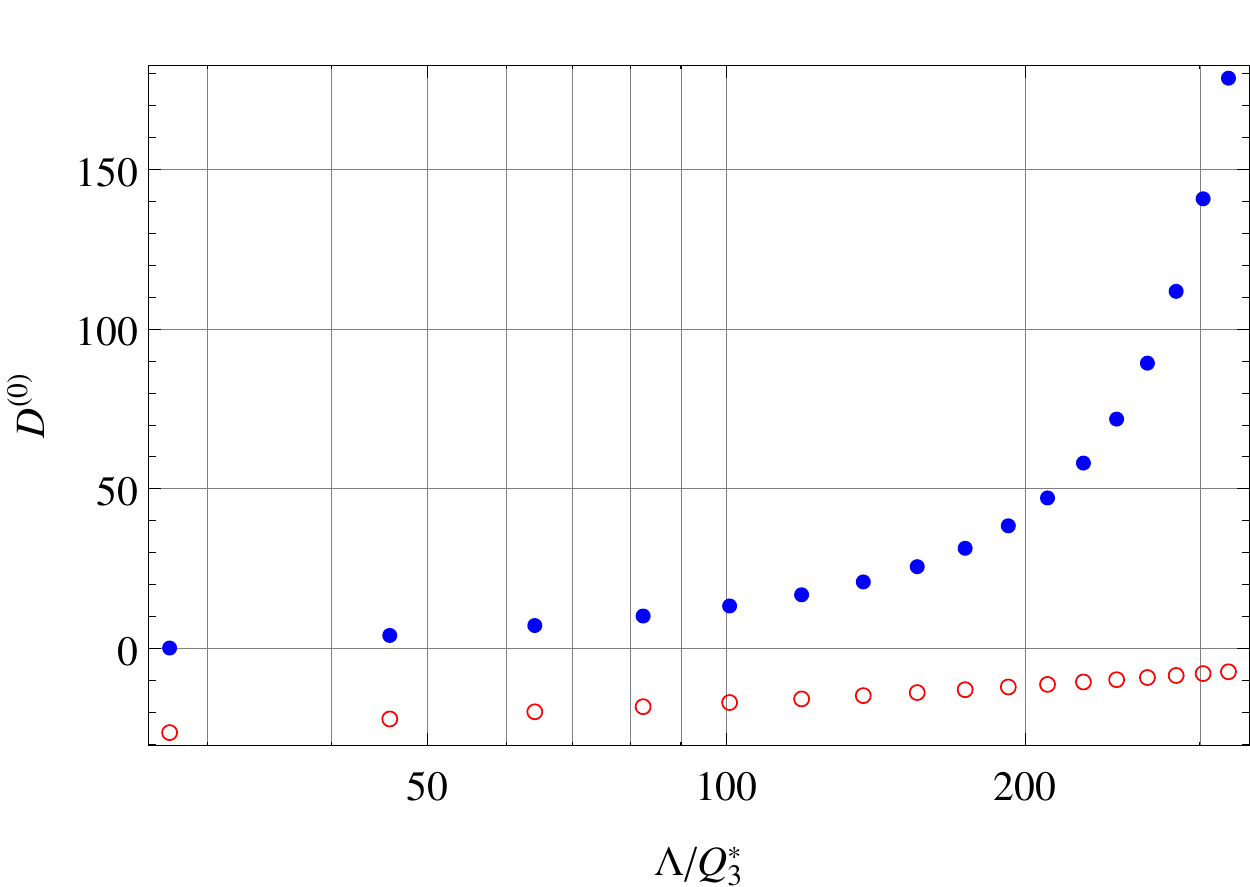}
\caption{\label{fig:LOD} (Color online)
  The dimensionless three-body low-energy constant 
  $D^{(0)}$ for various cutoff values $\Lambda$,
  in units of the binding momentum of the trimer excited state $Q_3^*$.
  The upper branch (solid blue circles) corresponds to fitting 
  the first unrenormalized excited state to the $^4$He trimer excited state
  of the PCKLJS potential; 
  the lower branch (open red circles) corresponds to
  fitting instead the second unrenormalized excited state.
}
\end{center}
\end{figure}

In the three-boson system, we demand that the binding energy $B_3^*$ 
of the trimer excited state be obtained at all values of the cutoff.
This can be achieved in different ways, depending on which unrenormalized
state we bring to the excited-state energy with the three-body force. 
Different unrenormalized states lead to different functions $D^{(0)}(\Lambda)$.
The resulting three-body LEC is plotted in Fig. \ref{fig:LOD} for various 
values of $\Lambda/Q_3^*$,
when the PCKLJS value of the binding energy of the trimer excited
state is used as input.
The two branches correspond to fitting the unrenormalized first and second
excited 
states to the excited trimer. 
Similar cutoff dependence arises from the LM2M2 potential.

The structure shown in Fig. \ref{fig:LOD} is not the same as
Eq. \eqref{limcyc}: Not only is the form different,
but also our $D^{(0)}(\Lambda)$ depends on the value of $Q_2$.
In contrast to the regularization procedure that led to Eq. \eqref{limcyc},
here we can choose which unrenormalized state to use.
Equation \eqref{limcyc} comes from studies of a non-local potential,
while we employ a local one
with the same cutoff in the two- and three-body sectors,
suggesting that this is the source of the differences.
However, since $D^{(0)}(\Lambda)$ is not an observable, it may
depend on the regularization scheme; only observables need
to be independent of the regularization scheme.
In the following we use the upper branch of Fig. \ref{fig:LOD}.

\begin{figure}
\begin{center}
\includegraphics[width=8.6 cm]{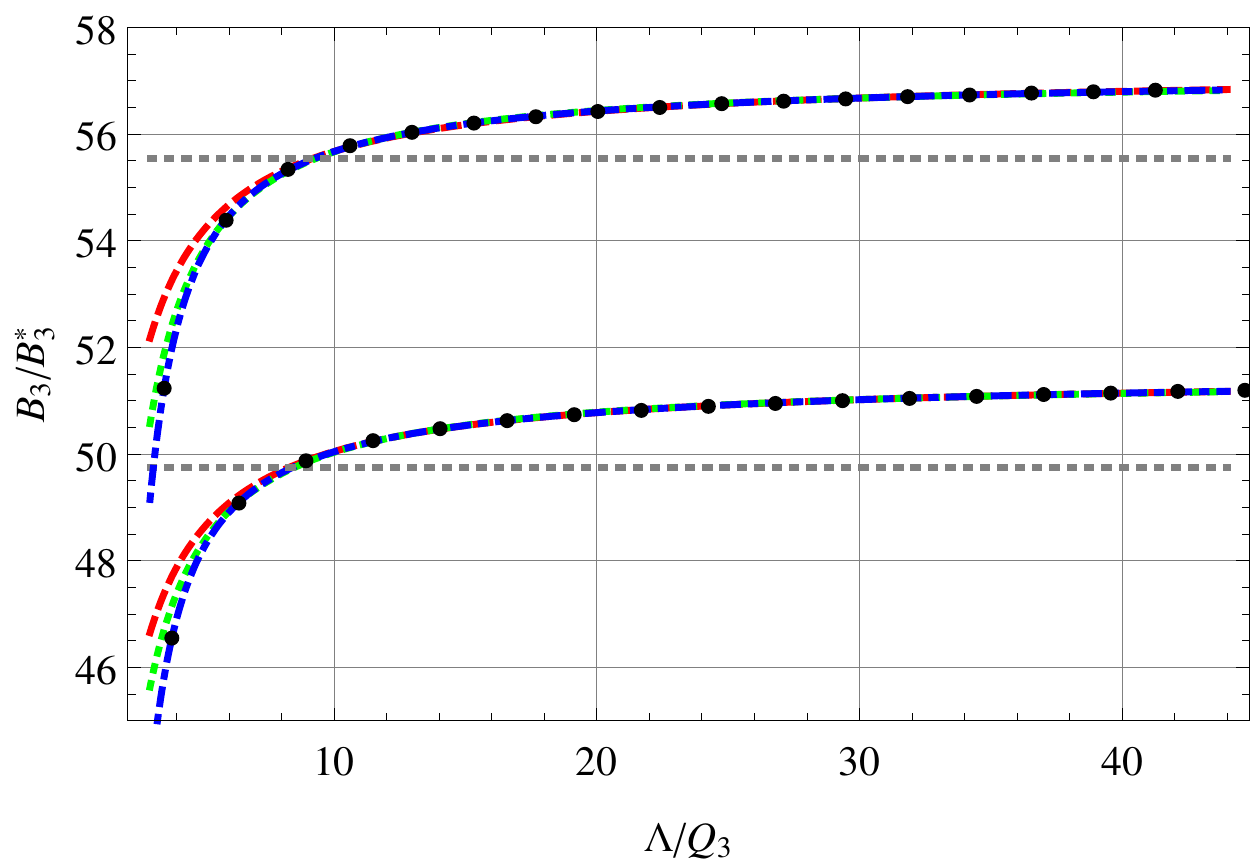}
\caption{\label{fig:HeE3} (Color online)
  The ratio of ground-to-excited-state binding energies of the 
  $^4$He trimer, $B_3/B_3^*$, 
  for various cutoff values $\Lambda$,
  in units of the asymptotic ground-state binding momentum $Q_3$.
  The upper and lower (black) circles correspond to the LO EFT calculation
  based on, respectively, LM2M2 and PCKLJS potentials.
  The horizontal (gray) dashed lines indicate the values calculated 
  directly from the corresponding potential \cite{HiyKam12a}.
  Also plotted are fits to powers of $Q_3/\Lambda$ in Eq. \eqref{B3fit} 
  cut after the first (red dashed lines), second (green dotted lines) and third
  (blue dot-dashed lines) terms.
}
\end{center}
\end{figure}

With the LECs fixed, we can predict the ground-state trimer energy.
The ratio of the binding energies of the trimer ground and excited states 
is plotted in Fig. \ref{fig:HeE3}.
The numerical results can be fitted by a series in
powers of the small parameter $Q_3/\Lambda$,
\begin{eqnarray}
\frac{B_3(\Lambda)}{B_3^*}&=&\frac{B_{3}(\infty)}{B_3^*}
\left[1 + \alpha_3 \frac{Q_3}{\Lambda} 
        + \beta_3 \left(\frac{Q_3}{\Lambda}\right)^2
        + \gamma_3 \left(\frac{Q_3}{\Lambda}\right)^3 
\right.
\nonumber\\
&& \left. \qquad\qquad
+ \ldots\right],
\label{B3fit}
\end{eqnarray}
where $Q_3$ itself is calculated from $B_{3}(\infty)$.
To check the stability of such a fit, we cut this series after each term
and fit to the calculated data. The resulting parameters 
are summarized in Table \ref{tbl:B3}. They have natural size
and similar values for the two potentials. 
The corresponding curves,
plotted in Fig. \ref{fig:HeE3},
are very close to each other for $\Lambda/Q_3$ beyond about 8.
This all suggests good
convergence to the zero-range limit.

\begin{table}
\begin{center}
\caption{Dimensionless parameters of the fit \eqref{B3fit} to 
  the trimer ground-state energy. The upper (lower) results correspond to the
  LM2M2- (PCKLJS-) based EFT.}
\label{tbl:B3}
\vspace{0.3cm}
{\renewcommand{\arraystretch}{1.25}%
\begin{tabular}
{c@{\hspace{5mm}} c@{\hspace{5mm}} c@{\hspace{5mm}} c}
\hline
\hline 
 $B_{3}(\infty)/B_3^*$ & $\alpha_3$ & $\beta_3$ & $\gamma_3$ \\ 
\hline 
 57.18           & $-0.26$    & ---     & ---      \\
 57.10           & $-0.21$    & $-0.39$ & ---      \\
 57.16           & $-0.26$    & 0.20    & $-2.04$  \\
 \hline 
 51.51           & $-0.28$    & ---      & ---     \\
 51.47           & $-0.25$    & $-0.26$  & ---     \\
 51.52           & $-0.31$    & 0.45     & $-2.66$ \\
\hline
\hline
\end{tabular}}
\end{center}
\end{table}

For the LM2M2 input, the asymptotic value of the binding energy is
$B_{3}(\infty)/B_3^*\simeq 57.15(4)$,
corresponding to $Q_3\simeq 0.045\, a_0^{-1}$, while
for the PCKLJS input, 
$B_{3}(\infty)/B_3^*\simeq 51.50(3)$,
corresponding to $Q_3\simeq 0.046\, a_0^{-1}$.
For the asymptotic value we took the mean of the various fits, and
the error given above reflects only their spread.
A naive estimate of the relative truncation error is 
$Q_3 r_2/2 \sim 0.3$, while cutoff variation gives about $0.1$,
because the magnitude of the dominant coefficient $\alpha_{3}$ is $\sim 1/3$.
For comparison, the ratio calculated directly from the potentials is
$55.53$ and $49.75$ for LM2M2 and PCKLJS, respectively \cite{HiyKam12a}.
This agreement at the 5\% level is somewhat accidental, since 
we have noticed that fitting the two-body LEC to
the scattering length $a_2$ instead of the binding energy $B_2$ results in
an agreement at the 20\% level instead.
In Table \ref{tbl:B3prime},
we list $B_{3}(\infty)/B_3^*$ for these potentials depending
on the input used, either $B_2$ or $a_2$.
Presumably, carrying out the EFT expansion around
the dimer pole in the complex momentum plane 
instead of the origin is better
because one starts closer to the positions of the poles representing
more-particle bound states \cite{Phillips:1999hh,Griesshammer:2004pe}.
Therefore, below we continue to use $B_2$ as input.
Since the two fitting procedures differ by NLO terms,
we assign a relative error of about 0.2 to our LO result,
which makes the extrapolation error completely negligible.

\begin{table}
\begin{center}
\caption{Asymptotic values of the trimer ground-state energy
  in units of the excited-trimer binding energy, depending on the
  two-atom input: dimer binding energy or scattering length.
  The upper (lower) results correspond to the
  LM2M2 (PCKLJS)-based EFT. 
  The error is only that which comes from the fitting procedure.
  Also, for comparison we list the values obtained \cite{HiyKam12a}
  directly from the corresponding potential.
  }
\label{tbl:B3prime}
\vspace{0.3cm}
{\renewcommand{\arraystretch}{1.25}%
\begin{tabular}
{c@{\hspace{5mm}} c}
\hline
\hline 
input & $B_{3}(\infty)/B_3^*$   \\ 
\hline 
 $B_2=1.3094$ mK  & $57.15(4)$ \\
 $a_2=100.23$ \AA & $65.30(3)$  \\
 direct \cite{HiyKam12a} & $55.53$ \\
\hline
 $B_2=1.6154$ mK  & $51.50(3)$ \\
 $a_2=90.42$ \AA  & $59.81(2)$ \\
 direct \cite{HiyKam12a} & $49.75$ \\
\hline
\hline
\end{tabular}}
\end{center}
\end{table}

As for two bosons, we can look at predictions for scattering as well.
To avoid dealing with continuum wave functions, we put our system in an
isotropic harmonic trap of frequency $\omega$
and calculate the two-body ($E_2$) and three-body ($E_3$) 
energies for various trapping frequencies.
As the trap is weakened, that is, $\omega$ becomes small,
the lowest two- and three-body states approach the free dimer and the 
free trimers,
while higher states form the respective scattering continua.
For the three-body system, the energies above the dimer energy
can be used to extract atom-dimer scattering parameters,
as long as the harmonic-oscillator length 
$a_{ho}=1/\sqrt{2\mu \omega}$, where $\mu\simeq 2m/3$
is the atom-dimer reduced mass, is larger than the dimer size $\sim a_2$. 
In this case, we can treat the dimer as a point-like particle, and
the solution for two-body
scattering inside a trap \cite{BusEtAl98,SteRotBar10} can be used to
extract the free-space scattering parameters.
For sufficiently small $E_3$, 
\be
\sqrt{2}l\frac{\Gamma[(3-\eta)/4]}{\Gamma[(1-\eta)/4]}\simeq
\frac{a_2}{a_3}\left(1- \frac{a_{3} r_{3}}{4 a_2^2} \eta l^2 \right),
\label{Etoscatttrap}
\ee
where 
$\eta=2 (E_3-E_2)/\omega$, $l=a_2/a_{ho}$, and $a_3$ and $r_3$ 
are the atom-dimer scattering length and effective range,
respectively.

\begin{figure}
\begin{center}
\includegraphics[width=8.6 cm]{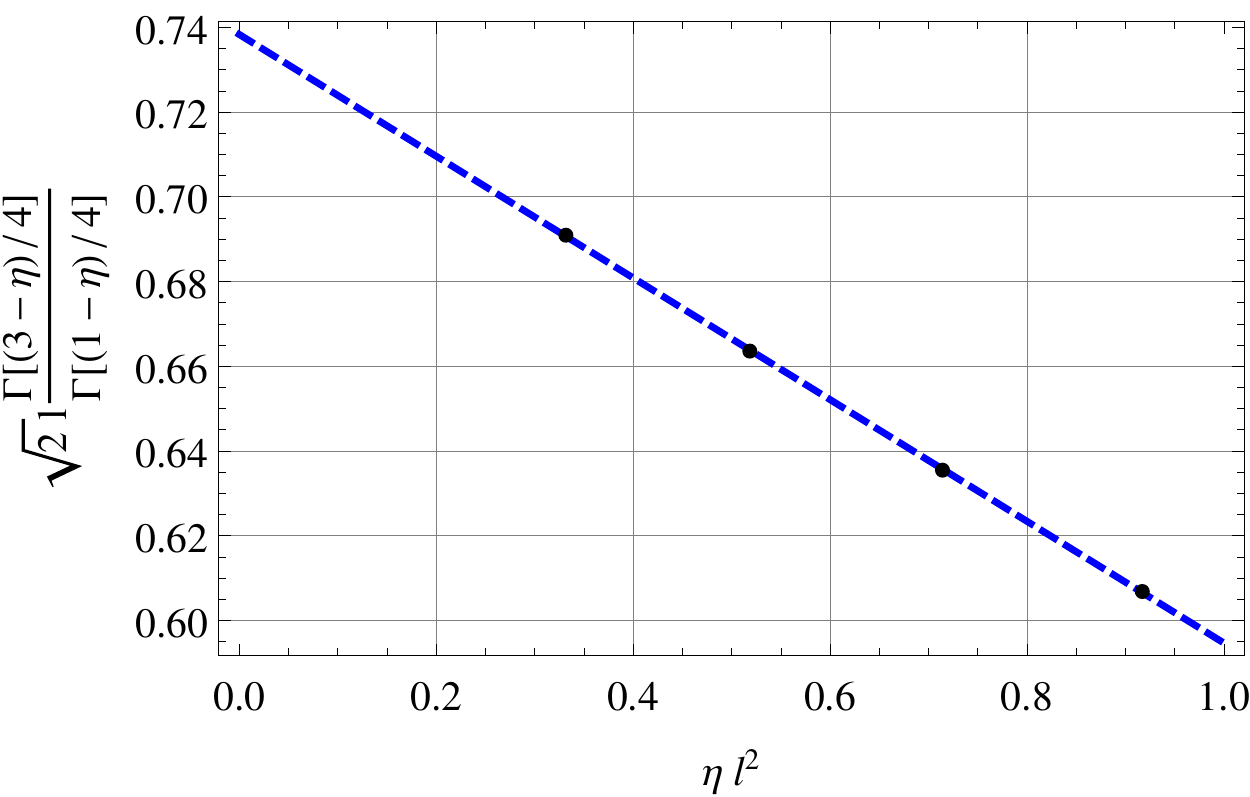}
\caption{\label{fig:omegaenergies} (Color online)
The left-hand side of Eq. \eqref{Etoscatttrap} as a function
of $\eta$, the difference between
three- and two-body energies in units of 
half the harmonic oscillator frequency,
multiplied by $l^2=(a_2/a_{ho})^2$, at a cutoff $\Lambda/Q_3^*=27.5$.
Also plotted is the fit with the right-hand side of Eq.  \eqref{Etoscatttrap}.
}
\end{center}
\end{figure}

\begin{figure}
\begin{center}
\includegraphics[width=8.6 cm]{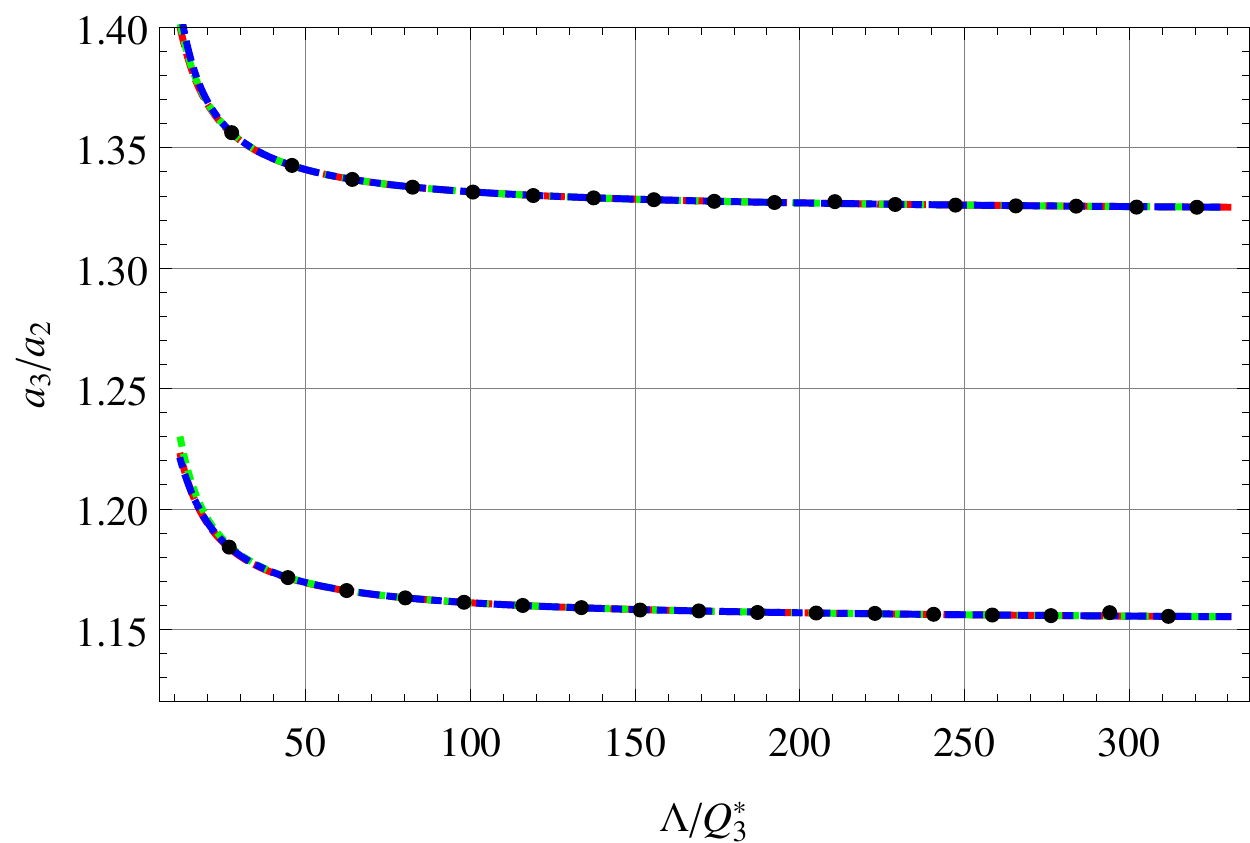}
\caption{\label{fig:HeAad} (Color online)
 The ratio of the atom-dimer scattering length $a_{3}$ to
 the two-atom scattering length $a_2$ for various cutoff values $\Lambda$
 in units of $Q_3^*$. The upper and lower (black) circles correspond to the
 LO EFT calculation based on, respectively, PCKLJS and LM2M2 potentials.
 Also plotted are fits to powers of $Q_3^*/\Lambda$ in Eq. \eqref{andfit}, 
 cut after the first (red dashed lines), second (green dotted lines) and third
 (blue dot-dashed lines) terms.
}
\end{center}
\end{figure}

Results for the left-hand side of Eq. \eqref{Etoscatttrap} at
a cutoff $\Lambda/Q_3^* = 27.5$ are given as a function of
$\eta l^2$ in Fig. \ref{fig:omegaenergies}.
Fitting them with the right-hand side of Eq. \eqref{Etoscatttrap}
allows us to extract $a_{3}$ and $r_{3}$ at that cutoff.
The resulting atom-dimer scattering length $a_{3}$ is shown in
Fig. \ref{fig:HeAad} for various cutoff values. 
We fit the cutoff dependence with
\beqa
\frac{a_{3}(\Lambda)}{a_{2}}&=&\frac{a_{3}(\infty)}{a_{2}}
\left[1 + \tilde\alpha_3 \frac{Q_3^*}{\Lambda} 
+ \tilde\beta_3 \left(\frac{Q_3^*}{\Lambda}\right)^2
+ \tilde\gamma_3 \left(\frac{Q_3^*}{\Lambda}\right)^3
\right.
\nonumber\\
&& \left. \qquad\qquad
+ \ldots\right].
\label{andfit}
\eeqa
The resulting parameters are summarized in Table \ref{tbl:and} 
and the corresponding curves plotted in Fig. \ref{fig:HeAad}.
Again, convergence to the zero-range limit is good.
The asymptotic value is
$a_{3}(\infty)=1.153(1)\, a_2$
($a_{3}(\infty)=1.322(1)\, a_2$) for the EFT based on the LM2M2 (PCKLJS) values.
An estimate of the relative truncation error is $Q_3^*r_2/2\sim 0.1$,
since our most important source of systematic error here is
the determination of the three-body force parameter.

\begin{table}
\begin{center}
\caption{Dimensionless parameters of the fit \eqref{andfit} to 
  the atom-dimer scattering length. The upper (lower) results correspond to the
  LM2M2- (PCKLJS-) based EFT.}
\label{tbl:and}
\vspace{0.3cm}
{\renewcommand{\arraystretch}{1.25}%
\begin{tabular}
{c@{\hspace{5mm}} c@{\hspace{5mm}} c@{\hspace{5mm}} c@{\hspace{5mm}} c}
\hline
\hline 
 $a_3(\infty)/a_2$ & $\tilde\alpha_3$ & $\tilde\beta_3$ & $\tilde\gamma_3$ \\ 
 \hline 
1.153  & 0.72 & ---   & ---     \\
1.153  & 0.70 & 1.03  & ---     \\
1.153  & 0.71 & 1.11  & $-14.2$ \\
\hline 
1.322  & 0.69 & ---     & ---   \\
1.322  & 0.69 & 0.21    & ---   \\
1.322  & 0.70 & $-0.54$ & 15.3  \\
\hline
\hline
\end{tabular}}
\end{center}
\end{table}

The atom-dimer scattering length was first 
shown by Efimov \cite{Efi79} to have a log-periodic structure
in $a_2\Lambda_\star$, a result which was reproduced in 
EFT \cite{BedHamKol99,BedHamKol99b}.
Several authors calculated this quantity using various $^4$He-$^4$He potentials.
In Table \ref{tbl:HeaAd} we summarize the available results for the
LM2M2 potential, and compare them with our asymptotic value.
The results from Ref. \cite{BluGre00}
are in (marginal) agreement with ours, considering our truncation error. 
We agree very well 
with the very precise results of Refs. \cite{Rou03,KolMotSan04,LazCar06,Del15},
and with the LO EFT result of Ref. \cite{BraHam03}.
Higher ERE parameters can be obtained similarly.

\begin{table}
\begin{center}
\caption{The $^4$He atom-dimer scattering length
in units of the atom-atom scattering length, as calculated with
the LM2M2 potential, and with LO EFT in Ref. \cite{BraHam03}
and in this work. 
The cited error for our result reflects only uncertainties
in the extrapolation procedure; for an extensive discussion
of systematic errors, see Sec. \ref{sec:results3}.
}
\label{tbl:HeaAd}
\vspace{0.3cm}
{\renewcommand{\arraystretch}{1.25}%
\begin{tabular}
{c@{\hspace{3mm}} c@{\hspace{3mm}} c@{\hspace{3mm}} c@{\hspace{3mm}} 
c@{\hspace{3mm}} c@{\hspace{3mm}} c@{\hspace{3mm}} c}
\hline
\hline 
Ref.&\cite{BluGre00}&\cite{KolMotSan04}&\cite{Rou03,LazCar06,Del15}
& \cite{BraHam03} & this work \\ 
\hline
$a_3/a_2$ & 1.26 & 1.152(5) & 1.151(2) & 1.128 & 1.153(1) \\
\hline
\hline
\end{tabular}}
\end{center}
\end{table}

At NLO, no new three-body force is necessary for renormalization 
\cite{BedHamKol99b}, and corrections linear in $r_2$
can be predicted \cite{Ji:2011qg}.
At N$^2$LO, a two-derivative three-body force appears \cite{Ji:2012nj}
and a second three-body input is needed.
Taking it to be $a_3$, Ref. \cite{Ji:2012nj} presents results
for the atom-dimer phase shifts and the trimer ground state.
Comparing with the LO results from Refs. \cite{BedHamKol99,BedHamKol99b},
reasonable convergence with order is found.

\subsection{Four, five and six bosons}
\label{sec:results456}

We have seen that a three-body counterterm is needed to stabilize the 
three-body system. Are more terms needed to stabilize heavier systems?

The answer for $N=4$ is believed to be known: 
no four-body counterterm is needed at LO.
This is a consequence of the 
apparent convergence with the cutoff of the calculated
tetramer energy, at least for one regulator function
and a certain cutoff range \cite{PlaHamMei04,Hammer:2006ct}.
It is manifested in a correlation between the tetramer and trimer
binding energies for fixed dimer energies but different values
of $\Lambda_\star$.
In nuclear physics, the equivalent correlation between the $^4$He and
$^3$H binding energies calculated with different nuclear potentials
is known as the Tjon line \cite{Tjo75}.

Pushing the number of particles up, do we need to add other counterterms?
In the nuclear case, where there are four-state fermions (protons 
and neutrons with spin up and down),
such a counterterm would involve at least two derivatives because
Pauli exclusion forbids more than four nucleons to come together.
Indeed, the binding energy of $^6$Li was calculated within a
similar EFT and seems to be converged without additional counterterms
\cite{SteBarKol07}.
However, is the same true also for bosons, where 
no-derivative contact counterterms exist? 

Here we calculate the tetramer, pentamer, and hexamer ground-state energies
at LO.
We show that these energies converge as the cutoff is increased,
dispensing with the need for higher-body interactions at this order.
We also construct generalized Tjon lines.

\begin{figure}
\begin{center}
\includegraphics[width=8.6 cm]{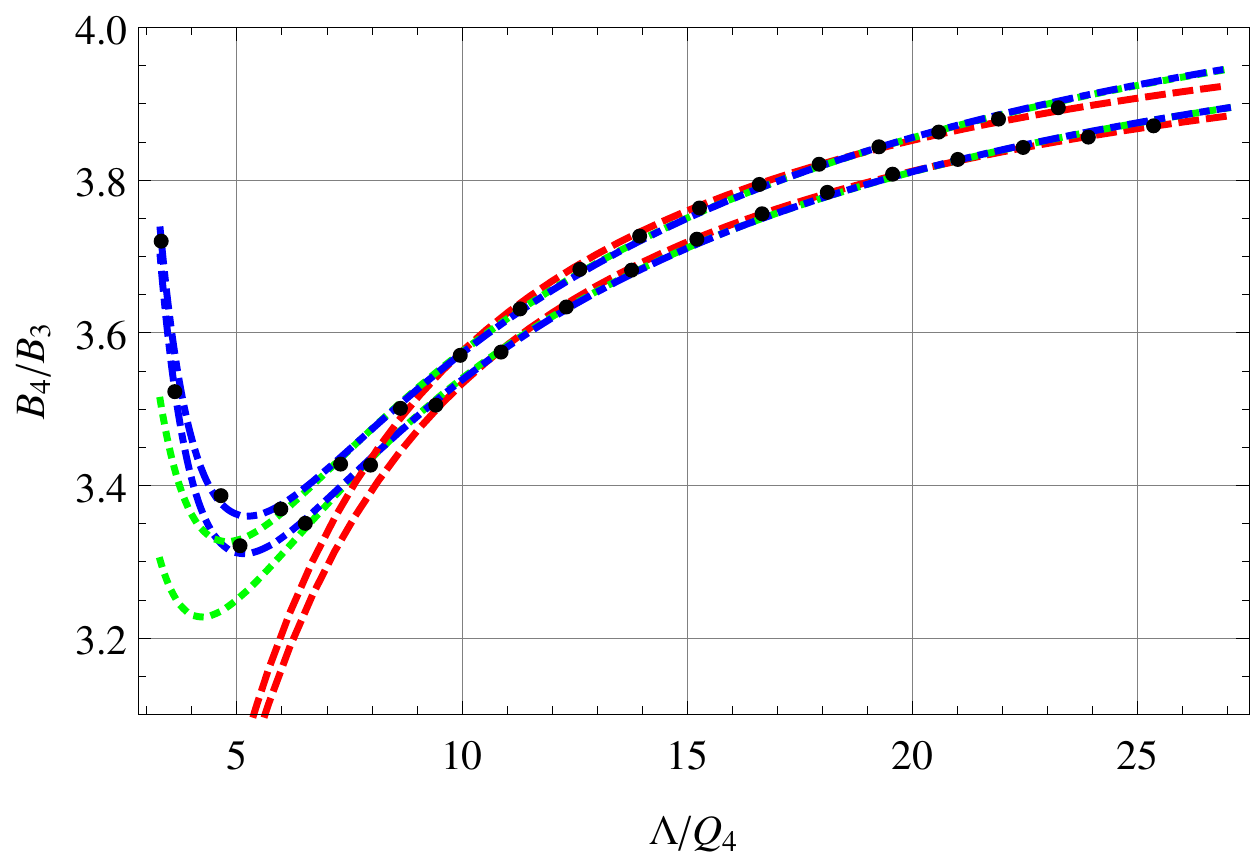}
\caption{\label{fig:HeE4} (Color online)
  The ratio of the ground-state energies of the $^4$He tetramer and 
  trimer, $B_4/B_3$, for various cutoff values $\Lambda$,
  in units of the asymptotic tetramer binding momentum $Q_4$.
  The upper and lower (black) circles correspond to the LO EFT calculation based
  on, respectively, LM2M2 and PCKLJS potentials.
  Also plotted are fits to powers of $Q_4/\Lambda$ in Eq. \eqref{BAfit} 
  cut after the first (red dashed line), second (green dotted) and third
  (blue dot-dashed) terms.
}
\end{center}
\end{figure}

\begin{figure}
\begin{center}
\includegraphics[width=8.6 cm]{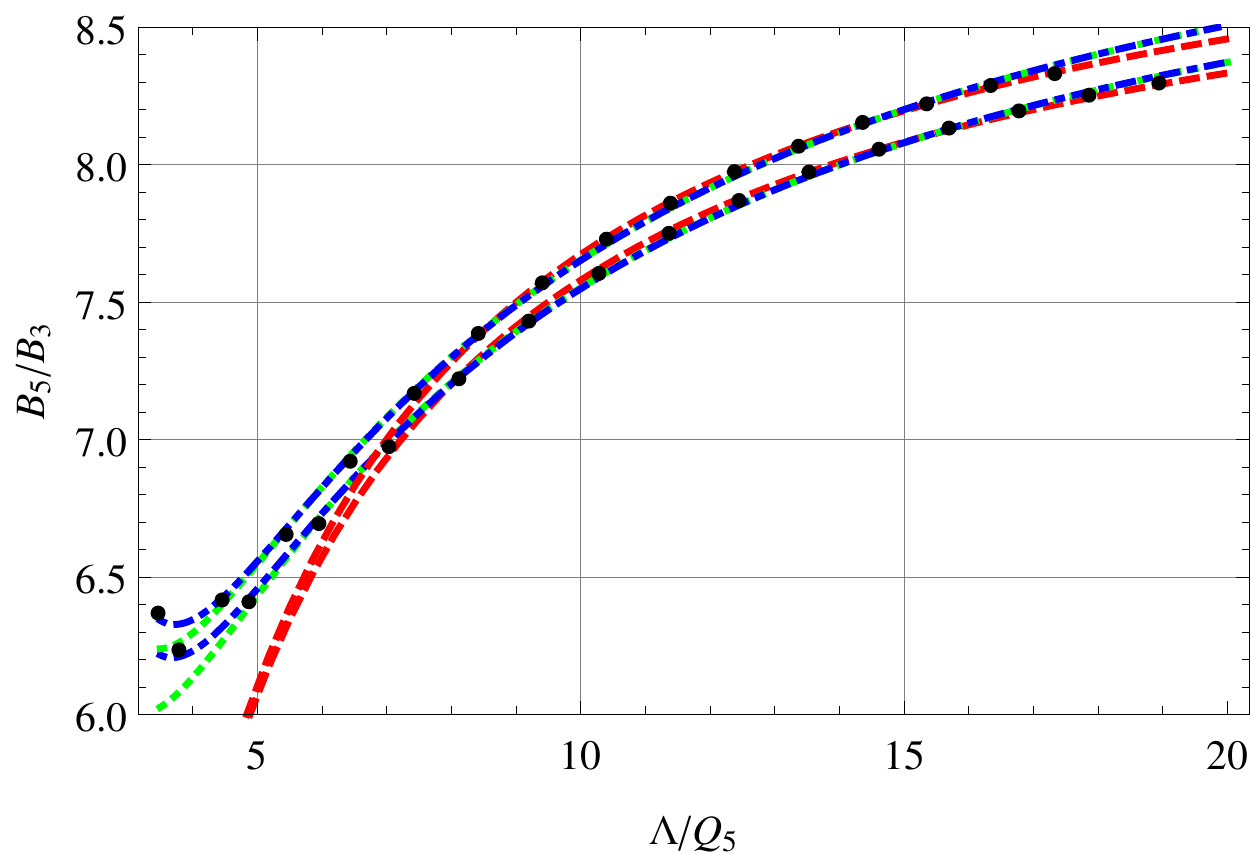}
\caption{\label{fig:HeE5} (Color online)
  The ratio of the ground-state energies of the $^4$He pentamer and 
  trimer, $B_5/B_3$, for various cutoff values $\Lambda$,
  in units of the asymptotic pentamer binding momentum $Q_5$.
  The upper and lower (black) circles correspond to the LO EFT calculation based
  on, respectively, LM2M2 and PCKLJS potentials.
  Also plotted are fits to powers of $Q_5/\Lambda$ in Eq. \eqref{BAfit} 
  cut after the first (red dashed lines), second (green dotted lines) and third
  (blue dot-dashed lines) terms.
}
\end{center}
\end{figure}

\begin{figure}
\begin{center}
\includegraphics[width=8.6 cm]{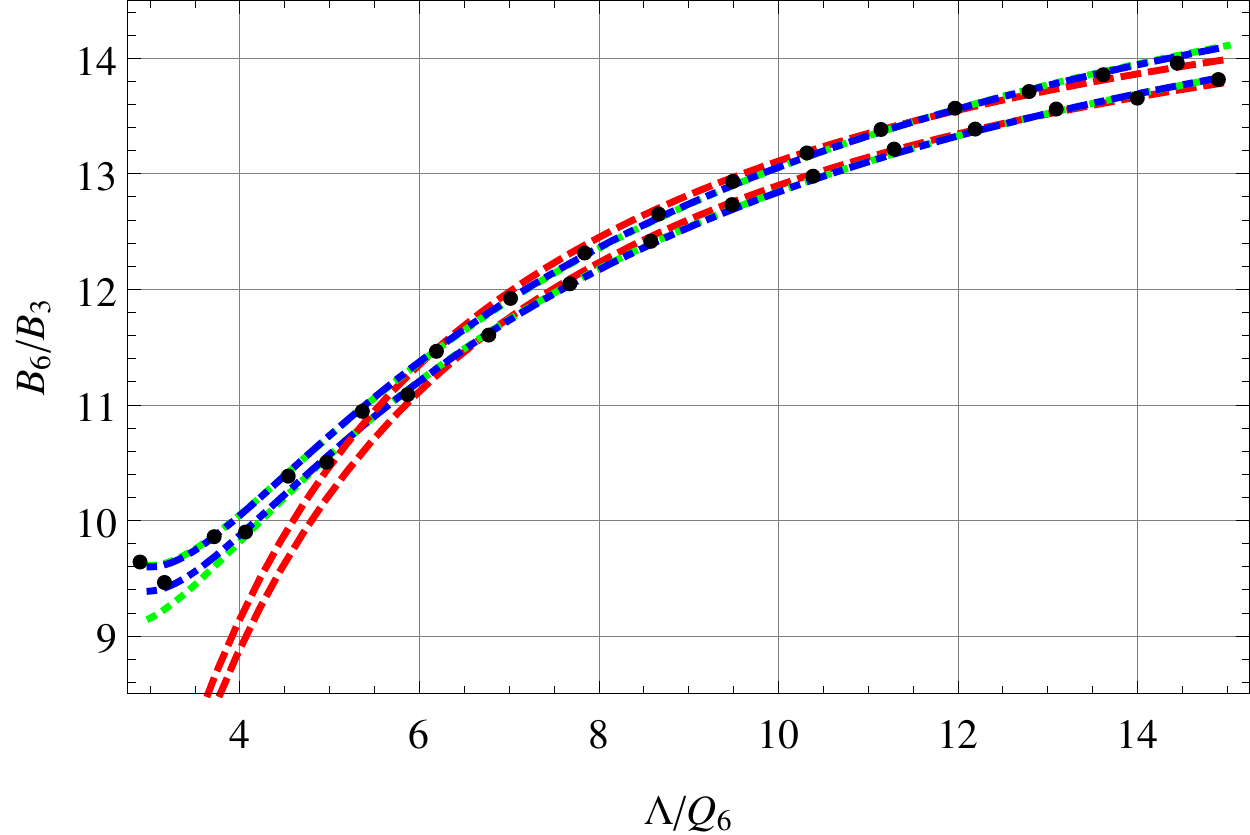}
\caption{\label{fig:HeE6} (Color online)
  The ratio of the ground-state energies of the $^4$He hexamer and 
  trimer, $B_6/B_3$, for various cutoff values $\Lambda$,
  in units of the asymptotic hexamer binding momentum $Q_6$.  
  The upper and lower (black) circles correspond to the LO EFT calculation based
  on, respectively, LM2M2 and PCKLJS potentials.
  Also plotted are fits to powers of $Q_6/\Lambda$ in Eq. \eqref{BAfit} 
  cut after the first (red dashed lines), second (green dotted lines) and third
  (blue dot-dashed lines) terms.
}
\end{center}
\end{figure}

The ratio of the tetramer and trimer ground-state energies is plotted
in Fig. \ref{fig:HeE4} for various cutoffs.
The corresponding plots for the pentamer and 
hexamer ground-state energies are presented in Figs. \ref{fig:HeE5} 
and \ref{fig:HeE6}, respectively.
Convergence is evident.
To strengthen the argument, we again fit the numerical results with
expressions of the type
\begin{eqnarray}
\frac{B_N(\Lambda)}{B_3(\Lambda)}&=&\frac{B_{N}(\infty)}{B_3(\infty)}
\left[1 + \alpha_N \frac{Q_N}{\Lambda} 
        + \beta_N \left(\frac{Q_N}{\Lambda}\right)^2
\right.
\nonumber\\
&& \left. \qquad\qquad
        + \gamma_N \left(\frac{Q_N}{\Lambda}\right)^3 
        + \dots\right],
\label{BAfit}
\end{eqnarray}
where $Q_N$ is calculated from $B_{N}(\infty)$ via Eq. \eqref{QA}.
The corresponding curves truncated at successive terms
are also shown in Figs. \ref{fig:HeE4}, \ref{fig:HeE5}, and  \ref{fig:HeE6}.

The fitting parameters for $N=4$ are shown in Table \ref{tbl:B4}.
The values are somewhat larger than for the ground-state trimer
in Table \ref{tbl:B3}, reflecting slower convergence.
But they still can be considered natural, 
suggesting that
the tetramer is likely within the EFT despite being considerably
more bound than the trimer.
Although we cannot fully exclude a mild $\ln \Lambda$ divergence,
adding such a term to Eq. \eqref{BAfit} requires a much smaller 
coefficient.
Thus we confirm, with a different
regulator than in Ref. \cite{PlaHamMei04,Hammer:2006ct},
that no four-body force is needed at LO.
Our results are qualitatively similar for $N=5,6$.
The fitting parameters $\alpha_{5,6}$, $\beta_{5,6}$, and $\gamma_{5,6}$
for the pentamer and the hexamer binding energies are, again, of natural size,
while an $\ln \Lambda$ term is much smaller.
The good convergence of the pentamer and hexamer ground-state energies
for our regulator is evidence
that no five- and six-body forces are needed at LO, either.

\begin{table}
\begin{center}
\caption{Dimensionless parameters of the fit \eqref{BAfit} to 
  the tetramer-to-trimer binding-energy ratio. 
  The upper (lower) results correspond to the
  LM2M2 (PCKLJS)-based EFT.}
\label{tbl:B4}
\vspace{0.3cm}
{\renewcommand{\arraystretch}{1.25}%
\begin{tabular}
{c@{\hspace{5mm}} c@{\hspace{5mm}} c@{\hspace{5mm}} c@{\hspace{5mm}} c}
\hline
\hline 
 $B_4(\infty)/B_3(\infty)$ & $\alpha_4$ & $\beta_4$ & $\gamma_4$ \\ 
 \hline 
4.128  & $-1.34$ & ---  & ---     \\
4.240  & $-2.06$ & 4.93 & ---     \\
4.238  & $-2.02$ & 4.06 & 4.30    \\
\hline 
4.090  & $-1.36$ & ---  & ---     \\
4.165  & $-1.90$ & 4.02 & ---     \\
4.157  & $-1.80$ & 2.32 & 7.99    \\
\hline
\hline
\end{tabular}}
\end{center}
\end{table}

Our converged values for the tetramer-to-trimer ratio of ground-state energies
are $B_{4}(\infty)/B_3(\infty)\simeq 4.20(6)$ and $4.14(4)$ 
for LM2M2 and PCKLJS potentials, respectively.
This is to be compared with the corresponding 
values predicted 
directly from these potentials, 
respectively 4.42 and 4.35 \cite{HiyKam12a}. 
Similarly, we find that our converged values 
for the pentamer-to-trimer ratio of ground-state energies 
are $B_{5}(\infty)/B_3(\infty)\simeq 9.5(2)$ and $9.3(2)$ 
for LM2M2 and PCKLJS, respectively.
For the hexamer-to-trimer ratio,
we find $B_{6}(\infty)/B_3(\infty)\simeq 16.3(5)$ 
for LM2M2 and $16.0(4)$ for PCKLJS.
We are not aware of five- nor six-body calculations using the PCKLJS
potential. 
The values predicted directly from the LM2M2 potential
are 10.33(1) for pentamer and 18.41(2) for hexamer \cite{BluGre00}.
In Table \ref{tbl:HeBE} we summarize the binding energies ratios
for $N\ge4$ systems, as well as available results calculated
directly from $^4$He-$^4$He potentials.

\begin{table}
\begin{center}
  \caption{The $N$-body $^4$He binding energies, in units of the trimer
    binding energy, for $N=4,5,6$.
    Our results are compared to those obtained with the
    PCKLJS \cite{HiyKam12a}, LM2M2 \cite{BluGre00}
    and TTY \cite{Lew97} potentials,
    as well as a soft-core potential \cite{GatKieViv11}.
    The cited errors for our results reflect only uncertainties
    in the extrapolation procedure; for an extensive discussion
    of systematic errors, see Sec. \ref{sec:results456}.
}
\label{tbl:HeBE}
\vspace{0.3cm}
{\renewcommand{\arraystretch}{1.25}%
\begin{tabular}
{c@{\hspace{3mm}}c@{\hspace{3mm}} c@{\hspace{3mm}} c@{\hspace{3mm}}c@{\hspace{3mm}} c}
\hline \hline 
Ref.&\cite{HiyKam12a}&\cite{BluGre00}&\cite{Lew97}&\cite{GatKieViv11}&this work\\ \hline
$B_4/B_3$ & 4.35 & 4.44(1)  & 4.49(2)   & 4.500  & 4.20(6) \\
$B_5/B_3$ & ---  & 10.33(1) & 10.519(8) & 10.495 & 9.5(2)  \\
$B_6/B_3$ & ---  & 18.41(2) & 18.50(2)  & 18.504 & 16.3(5) \\
\hline
\hline
\end{tabular}}
\end{center}
\end{table}

The errors reported above are only fitting errors. 
Our naive systematic error is much larger,
ranging from 55\% for $N=4$ to 90\% for $N=6$, if we use
$Q_Nr_2/2\sim (r_2/2a_2)\sqrt{2B_N/NB_2}$ 
for an estimate. 
Cutoff variation gives similar estimates, since the fitting
parameters are ${\cal O}(1)$.
Yet, compared with values obtained directly from the LM2M2 
potential \cite{BluGre00}, our central values for energy ratios are off by 
only about 5\% for $N=4$ and 15\% for $N=6$. 
The soft-core potential of Refs. \cite{GatKieViv11,Gattobigio:2012tk}
resembles our LO supplemented by some higher-order corrections,
so we might expect that NLO will supply most of the
difference between our results and those of Ref. \cite{BluGre00}.
This suggests that $Q_Nr_2/2$, with $Q_N$ defined by Eq. \eqref{QA}, 
is an overestimate of the systematic error.
Note that the spread among potential models,
which differ in the details of short-range physics,
is only about 2\%. It seems that, despite the increasing binding,
most of the dynamics of these atomic droplets takes
place at distances larger than $r_{\rm vdW}$,
with $1/Q_N$ an underestimate of the relevant distance.

\begin{figure}[t!]
\begin{center}
\includegraphics[width=8.6 cm]{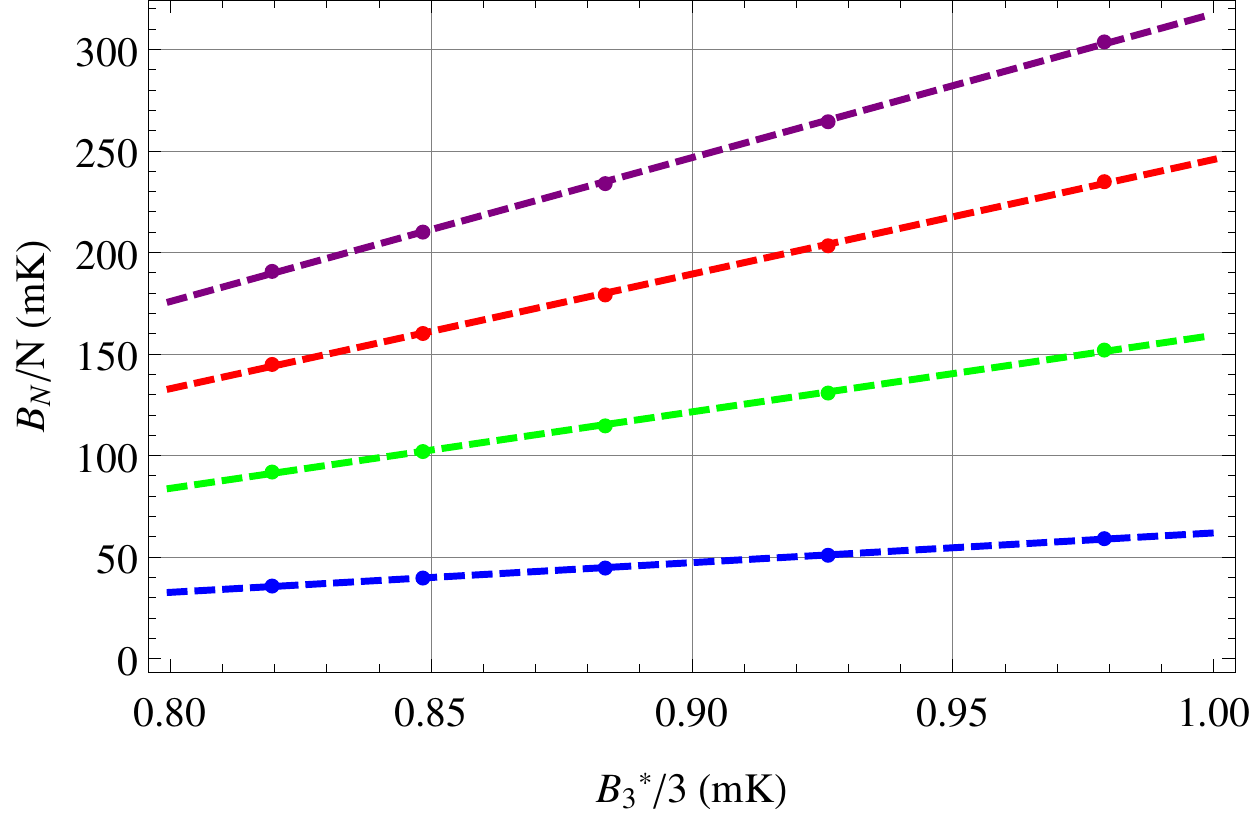}
\caption{\label{fig:GTjon} (Color online)
  Generalized Tjon lines: binding energies per particle, $B_N/N$,
  in mK
  for the $N=3$ (blue, bottom), $4$ (green, second from bottom), 
  $5$ (red, third from bottom) 
  and $6$ (purple, top) ground states for various values of the trimer
  excited-state energy per particle, $B_3^*/3$, in mK, at fixed
  dimer binding energy $B_2$.
  Here we used $B_2=1.6154$ mK and $\Lambda \simeq 1.22$ \AA$^{-1}$.
  Linear fits are also shown.
}
\end{center}
\end{figure}

The absence of higher-body forces at LO means
that, for fixed two-body input, the LO energies
of not only the ground tetramer, but also of the ground pentamer
and hexamer are determined by the three-body parameter.
Since we use the excited trimer energy as input, 
the same is true, as we have seen in Sec. \ref{sec:results3},
for the ground trimer. As a consequence, these energies
are all correlated.
Next we vary $Q_3^*$ at fixed $B_2$, therefore changing $B_3^*$, and
calculate $B_3$, $B_4$, $B_5$, and $B_6$. 
The generalized Tjon lines are plotted in
Fig. \ref{fig:GTjon}, 
where the correlation
between the ground-state binding energies $B_N/N$ 
and $B_3^*/3$ is evident.
The results presented in Fig. \ref{fig:GTjon} were obtained 
for a dimer binding energy $B_2=1.6154$ mK
and a cutoff $\Lambda \simeq 1.22$ \AA$^{-1}$.
Qualitatively similar results are obtained for higher cutoffs.
In the region we focus on, the generalized Tjon lines are approximately
straight with a derivative that increases monotonically
with $N$.
The corresponding linear fits are also shown in Fig. \ref{fig:GTjon}.

Changing $B_2$, line positions change,
but not, of course, the fact that a correlation exists among the
various energies. 
On dimensional grounds we can write
\be
\frac{B_N}{N}= c_N(B_2/B_3^*) \; \frac{B_3^*}{3}, 
\ee
where the $c_N(B_3^*/B_2)$ are universal, dimensionless functions of the only 
dimensionless ratio at LO, $B_3^*/B_2$. 
(At finite cutoff, $c_N$ depends also on $Q_3^*/\Lambda$.)
Results from phenomenological
two-body potentials with similar dimer binding energies
should fall on or close to these lines.
Because the scattering length of the $^4$He dimer
is so large, this system is relatively close to
the unitarity limit in the sense that the
first Efimov excited state exists.
The generalized Tjon lines at unitarity, while not exactly the same as 
those in 
Fig. \ref{fig:GTjon}, are not very different.
Expanding in a series in $B_2/B_3^*$,
\be
\frac{B_N}{N}= c_N(0) \; \frac{B_3^*}{3} 
+\ldots,
\label{BN/Nfit}
\ee
in terms of the universal set of numbers $c_N(0)$.
Eliminating $B_3^*$ in favor of $B_3$,
\be
\frac{B_N}{B_3}\approx \frac{Nc_N(0)}{3c_3(0)} \approx (N-2)^2,
\label{regular}
\ee
where the last approximation summarizes
our converged values in Table \ref{tbl:HeBE}.
Since $B_2/B_3\approx 10^{-2}$, 
Eq. \eqref{regular} applies to $N=2$ as well.

Near unitarity, then, the physics of few-boson clusters 
is essentially determined by the single parameter, $\Lambda_\star$,
that fixes $B_3$.
Moreover, the $N$ dependence is particularly simple:
$B_N/B_3$ seems to grow as $N^2$ as noticed in Ref. \cite{Nicholson:2012zp}.
In fact, our $N\le 6$ results, even though not
quite at unitarity, are well described by the empirical relation
\cite{Gattobigio:2013yda}
\be
\frac{B_N}{B_3}\approx \left[(N-3)\sqrt{\frac{B_4}{B_3}}+4-N\right]^2,
\label{regularmore}
\ee
which reduces to $(N-2)^2$ for $B_4/B_3\approx 4$.
Our results in Table \ref{tbl:HeBE} differ by less than 15\% from 
the $N\le 6$ ground-state ratios at unitarity 
obtained with a particular potential \cite{Ste10,Ste11}.
Note that the latter ratios are more consistent with 
\be
(N-2)^2\to \frac{N(N-1)(N-2)}{6}, 
\nonumber
\label{altenergies}
\ee
in Eq. \eqref{regular} and we cannot exclude such form.
However, the $7\le N\le 10$ values 
for the ratio $B_N/B_3$ in Ref. \cite{Ste10} 
are in much better agreement with $(N-2)^2$, and
for $N\ge 11$ energies grow even slower.
Ground-state ratios calculated from other potentials 
\cite{Kievsky:2014dua,YanBlu15} show different trends
as $N$ increases, but tend to agree with 
Eq. \eqref{regularmore}.
In EFT, saturation arises from a balance
between two- and three-body forces,
and the latter cannot be separated from the 
high-momentum components of the former in a cutoff-independent way.
In this context there is no obvious argument to justify a dependence on
$N$ of one form or another.
An EFT calculation along the lines of this paper
but for a wider range of $N$ values would be highly desirable.

\section{Conclusion} 
\label{sec:summary}

Summing up, we constructed an effective field theory for the few-boson system.
We have solved the leading-order Schr\"odinger equation for up to six 
particles using a correlated Gaussian basis and the stochastic variational 
method.
We have shown that various observables converge as the arbitrary 
ultraviolet cutoff increases.
The extrapolation to the infinite-cutoff limit corresponds
to a zero-range interaction.
It is therefore safe to assume that the dominant features
of bosonic systems, when within the domain of validity of EFT,
are described by two parameters: the coefficients
of the contact two- and three-body forces, which encode
the two-body scattering length $a_2$ (or the dimer binding energy $B_2$)
and a three-body parameter $\Lambda_\star$ (or a trimer binding energy).
Generalized Tjon lines were introduced, showing the correlation
among the ground-state energies of three, four, five and six particles
when $\Lambda_\star$ is varied at fixed $a_2$.

For concreteness, we have presented results for $^4$He systems.
As input, we could have used the experimental values
for the binding energies of the dimer, $B_2$,
and of the excited trimer, $B_3^*$. 
In order to gauge the convergence of the EFT, we opted instead
to use values from two modern $^4$He-$^4$He potentials,
LM2M2 and PCKLJS. 
These are representative of the experimental data, and therefore our numbers 
for other observables can be seen as EFT predictions.
At the same time, they allow us to compare our results with few-body energies
obtained directly from these potentials.
The latter energies contain information not only about $B_2$
and $B_3^*$, but also about a host of other $^4$He properties.
The point of an EFT is that this extra information
is only relevant in a marginal way, as we proceed to higher orders.
This is, of course, true only for observables 
for which the EFT expansion converges. 

Our results for $^4$He systems are in surprisingly good agreement with 
the literature.
They reproduce the LO results of Refs. \cite{Ji:2012nj} and 
\cite{PlaHamMei04} for $N=3$ and $4$, respectively,
and extend LO contact EFT to $N=5,6$.
The ground-state energies we calculate agree with direct results
from the same potential at a level that
ranges from about 5\% at $N=3$ to 15\% at $N=6$.
In contrast, a naive estimate based on the range of the potential
and Eq. \eqref{QA} as an estimate of the particle binding momentum
gives much larger errors. 
If not accidental, this agreement would imply that the range of applicability
of the EFT is much wider than expected.
Such optimism has become part of EFT folklore since a good description
was obtained in nuclear physics
for the triton \cite{Bedaque:1999ve} and, especially, the relatively
tight $alpha$ particle \cite{Platter:2004zs},
even though $^6$Li \cite{SteBarKol07}
did not come out particularly well despite having a similar binding energy
per particle.
Here we have added circumstantial evidence that the EFT expansion
works well beyond $N=3$ $^4$He atoms.
Systems within the range of applicability of
contact EFT share the same dynamical origin
as the three-body states that are usually labeled ``Efimov states.''

Note that the PCKLJS potential gives about 30\% more binding energy
for the dimer compared to LM2M2. 
Heavier systems are also more bound for PCKLJS,
but the excited and ground-state trimers are
more bound by just 15\% and 5\%,
respectively. Our $N=4,5,6$ ground-state energies 
for these potentials also differ by 5\% or less.
It seems that, once $B_3$ is obtained correctly, $B_{4,5,6}$ follow.
This pattern is consistent with 
potentials having similar values of $\Lambda_\star$, which determine 
the larger $B_{3,4,5,6}$, but 
somewhat different values of the large
$a_2$, which affect primarily the smaller $B_2$ and $B_3^*$. 
Perhaps the LO error in $B_3$, which we estimated at 20\%, is 
also a reasonable estimate of the error in $B_{4,5,6}$.

While we have presented results for $^4$He systems,
qualitatively similar results are obtained for other systems
with large scattering lengths, including the unitarity limit $1/a_2=0$.
The EFT captures Efimov physics at LO, where $\Lambda_\star$ 
determines the position of the geometric ladder of three-body bound states.
In systems with more particles, $\Lambda_\star$ also sets the 
scale for the binding energies.
For $N\le 6$, energies have the simple approximate scaling in 
Eq. \eqref{regular} or a similar form.
That $Q_3r_2/2$ might be a more realistic error
estimate than $Q_Nr_2/2$ is consistent with
the observation in Ref. \cite{YanBlu15} that the average interparticle
distance at unitarity is more or less constant with $N$,
and of ${\cal O}(1/Q_3)$.

Thus an EFT expanded around the zero-range limit 
captures the essence of universal bosonic systems.
Corrections to the zero-range limit, such as the two-body effective range
$r_2$, appear at next-to-leading order. 
Stronger arguments than presented here
about the convergence of the EFT expansion require higher-order
calculations. We plan to extend our calculation
to NLO in a future publication.

\section*{Acknowledgments}
We acknowledge useful discussions with N. Barnea, T. Frederico,
and H.-W. Hammer.
B.B. and U.v.K. 
thank the Institute for Nuclear Theory at the University of Washington 
for its hospitality during the Program INT-16-1 
``Nuclear Physics from Lattice QCD,'' when part of this work was carried out.
This material is based upon work supported in part
by a Chateaubriand Fellowship of the French Embassy in Israel (B.B.),
by the Pazi Fund (M.E.),  
by France's Centre National de la Recherche Scientifique under
a grant of IN2P3's Comit\'e de Financement des Projets de Physique Th\'eorique
(U.v.K.), and
by the U.S. Department of Energy, Office of Science, Office of Nuclear Physics,
under Award no. DE-FG02-04ER41338 (U.v.K.).


\end{document}